\documentclass{article}

\usepackage{arxiv}

\usepackage[utf8]{inputenc} 
\usepackage[T1]{fontenc}    
\usepackage{url}            
\usepackage{booktabs}       
\usepackage{amsfonts}       
\usepackage{nicefrac}       
\usepackage{microtype}      
\usepackage{lipsum}		
\usepackage{graphicx}
\usepackage{natbib}
\usepackage{doi}
\usepackage{newtxtext}
\usepackage{newtxmath}
\usepackage{authblk}

\title{Leveraging three-dimensionality for navigation in bluff-body wakes}

\author[1]{Vedasri Godavarthi$^1$\thanks{Corresponding author: vedasrig@g.ucla.edu}}
\author[2]{Kartik Krishna}
\author[2]{Steven L. Brunton}
\author[1]{Kunihiko Taira}
\affil[1]{Department of Mechanical and Aerospace Engineering, University of California, Los Angeles, CA 90095, USA}
\affil[2]{Department of Mechanical Engineering, University of Washington, Seattle, WA 98195, USA}

\hypersetup{
pdftitle={A template for the arxiv style},
pdfsubject={q-bio.NC, q-bio.QM},
pdfauthor={David S.~Hippocampus, Elias D.~Striatum},
pdfkeywords={First keyword, Second keyword, More},
}

\begin{document}
\maketitle

\begin{abstract}
	Biological flyers and swimmers navigate in unsteady wake flows using limited sensory abilities and actuation energies. Understanding how vortical structures can be leveraged for energy-efficient navigation in unsteady flows is beneficial in developing autonomous navigation for small-scale aerial and marine vehicles. Such vehicles are typically operated with constrained onboard actuation and sensing capabilities, making energy-efficient trajectory planning critically important. This study finds that trajectory planners can leverage three-dimensionality appearing in a complex unsteady wake for efficient navigation using limited flowfield information. This is revealed with comprehensive investigations by finite-horizon model-predictive control for trajectory planning of a swimmer behind a cylinder wake at $Re=300$. The navigation performance of three-dimensional (3D) cases is compared to scenarios in a two-dimensional (2D) wake. The underactuated swimmer is able to reach the target by leveraging the background flow when the prediction horizon exceeds one-tenth of the wake-shedding period, demonstrating that navigation is feasible with limited information about the flowfield. Further, we identify that the swimmer can leverage the secondary transverse vortical structures to reach the target faster than is achievable navigating in a 2D wake.
\end{abstract}


	\section{Introduction}
	\label{sec:intro}
	
	Small-scale aerial and marine vehicles are gaining traction in applications typically considered high-risk for manned vehicles. For these operations, vehicles need to traverse in unsteady flows, such as flying in urban environments \citep{watkins2020ten}, in the wake of artificial and natural structures, navigating behind large marine vehicles for maritime operations \citep{shukla2019ship}. Such wake flows are associated with strong unsteady 3D vortical disturbances. Autonomous vehicles often experience adverse effects when encountering such flows without prior information \citep{zereik2018challenges}. 

There have been studies on the effect of wake flows on the stability and performance of biological flyers and swimmers. The effect of 3D vortical perturbations generated by unsteady bluff body wakes on the aerodynamic coefficients and moments of insect flight and hummingbird flight show variable responses depending on the orientation and relative strength of the vortical perturbations \citep{ortega2014into,shyy2016aerodynamics}. The effect of three-dimensionality on locomotion is also studied in fish swimming \citep{liao2007review,lauder2015fish,maia2015streamwise} where fish undulate relative to the vortical structures for energy-efficient navigation.  These studies have been conducted at high speeds and also consider the interaction between the swimmer/flyer with the flowfield. Understanding how the 3D flow structures can be leveraged for navigation offers insights into biological locomotion and the development of autonomous navigation strategies. Hence, we consider trajectory planning for laminar 3D wakes.

Over recent years, there has been extensive research on trajectory-planning strategies for the navigation of autonomous vehicles in complex background flows. Given complete knowledge of the underlying environment, optimization methods such as graph-based algorithms \citep{kularatne2016time}, and stochastic optimization \citep{subramani2016energy} have been used to obtain optimal trajectories. Recently, reinforcement learning has been utilized for navigation in unsteady flows using limited local information \citep{colabrese2017flow,gunnarson2021learning,jiao2021learning}. 
Additionally, the Lagrangian coherent structures (LCS) theory has been used for energy-efficient path planning  \citep{senatore2008fuel,krishna2022finite} proposed a finite-horizon model predictive control approach to identify energy-efficient trajectories and their connection to LCS for navigating unsteady flows. 

In this study, we investigate the effect of three-dimensionality on navigation in unsteady wakes using the finite-horizon model predictive control approach used in \cite{krishna2022finite}.
We consider underactuated point swimmers to navigate in the 3D wake of a cylinder at a Reynolds number of 300, where the flow exhibits three-dimensionality due to secondary instabilities and compare their performance when navigating in 2D wakes. We reveal that secondary vortices in 3D flow facilitate faster navigation with limited information. The paper is outlined as follows. The computational setup and the flow physics of the cylinder flow are introduced in section~\ref{sec:comput_setup}. The finite-horizon model-predictive control approach is discussed in section~\ref{sec:MPC}. Section~\ref{sec:results} provides the results for trajectory planning in cylinder wakes. The conclusions are given in section~\ref{sec:conclusions}.

 \section{Computational setup}
 \label{sec:comput_setup}
We consider a 3D incompressible wake behind a circular cylinder obtained from a direct numerical simulation (DNS) at a diameter ($D$) based on Reynolds number, $Re=U_\infty D/\nu$ of 300, where $U_\infty$ and $\nu$ are freestream velocity and kinematic viscosity, respectively. The simulation is performed using the incompressible flow solver \textit{Cliff}, (Cascade Technologies Inc.) based on a second-order accurate finite volume method for spatial discretization and a fractional-step method for time stepping \citep{ham2004energy,ham2006accurate}. The computational domain extends to $-20\leq x/D \leq 30$ in the streamwise direction. The transverse extent is $-40\leq y/D \leq 40$ and the spanwise extent is $0 \leq z/D \leq 4$ to capture the wavenumber of secondary instabilities with a uniform discretization of 80 grid cells and a timestep of $\Delta t = 0.005.$  Hybrid grids are used with a structured mesh close to the cylinder and an unstructured mesh in the farfield region. We also obtain a 2D flow over a cylinder at the same Reynolds number using the same domain in the $x,\,y$ directions. It amounts to about 0.1 and 8.3 million cells for both 2D and 3D wakes, respectively. Further details on the computational setup can be found in \cite{kim2024influence}. For finite-horizon trajectory planning and visualization, we choose a subdomain $(x/D,\,y/D,\,z/D)\in [-2,10] \times [-2,2] \times [0,\,2].$

\section{Finite-horizon model predictive control}
\label{sec:MPC}
We use a finite-horizon model predictive control (MPC) approach developed by \cite{krishna2022finite} to perform trajectory optimization of a point swimmer. The swimmer dynamics is modeled as 
\begin{equation}
    \dot{\boldsymbol{x}}(t) = \boldsymbol{v}(\boldsymbol{x}(t),t)+\boldsymbol{u}(t),
    \vspace{-0.5em}
\end{equation}
where $\boldsymbol{x},\,\boldsymbol{u} \in \mathbb{R}^{n}$ are the position vector and actuation velocity of the swimmer while $\boldsymbol{v}$ is background flow velocity and $n=2,\,3$ for 2D and 3D flows. The background flow velocity is obtained from the DNS of the cylinder wake. When the actuation velocity is zero, the swimmer acts as a passive drifter. The swimmer dynamics is numerically integrated using a time step of $\Delta t = 0.05$. The actuation velocity is determined from the MPC optimization of the cost function given by
\begin{equation}
    J = \int\limits_{t_0}^{t_0+T_H}[\boldsymbol{e}(\tau)^T\mathbf{Q}\boldsymbol{e}(\tau)+\boldsymbol{u}(\tau)^T\mathbf{R}\boldsymbol{u}(\tau)]d\tau
     \vspace{-0.5em}
\end{equation}
subject to constraints on the component-wise actuation velocity with $|u_i(t)|\leq \eta_i$, where $\eta_i$ is the actuation velocity bounds on $i^{\rm th}$ velocity component. Here, $T_H$ is the time horizon over which the cost function is minimized, and $\boldsymbol{e}$ is the error of the current state from a target state, $\boldsymbol{x}(t)$, $\boldsymbol{e}(t) =\boldsymbol{x}(t)-\boldsymbol{x}_{\rm target}$. The matrix $\mathbf{Q}\in \mathbb{R}^{n\times n}$ is positive semi-definite that penalizes the state error throughout the trajectory, and $\mathbf{R}\in \mathbb{R}^{n\times n}$ is a positive definite matrix penalizing the actuation effort. Here the actuation velocity of the swimmer is penalized, although the acceleration could also be optimized.

For the current problem, $\mathbf{Q}$ is set to the identity matrix. When the actuation velocity is large i.e., $|u| \geq U_\infty$ (freestream), the trajectory reaches the target directly. We consider underactuated scenarios to see how the swimmer exploits the background flow. The local background velocity for the cylinder wake varies in the streamwise, transverse, and spanwise ($x,\,y,\,z$) directions, with the streamwise velocity being dominant.  Thus, the actuation efforts are penalized differently in different directions, and $\mathbf{R}$ is a diagonal matrix where the $i^{\rm th}$ diagonal element is given as $R_{ii}=\gamma_i$ and $\gamma_i$ is the actuation penalization in $i^{\rm th}$ direction. The time horizon $T_H$ is another key parameter, as larger $T_H$  requires larger predictive capabilities i.e., more temporal information about the background flow to the swimmer. 

\section{Navigating three-dimensional wake flows}
\label{sec:results}
\begin{figure}
    \centering
    \includegraphics[scale=0.9]{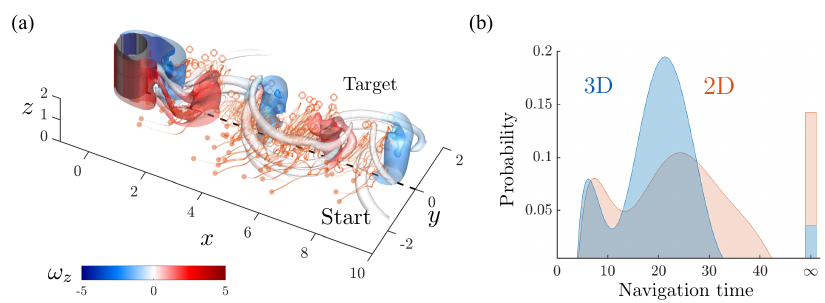}
    \caption{Trajectories of swimmers crossing the wake for 3D flow over a cylinder at $Re=300$ visualized using iso-surface of $Q-$criterion $Q=0.5$, colored by spanwise vorticity $\omega_z$ in (a) isometric view. (b) Probability distribution of total navigation time in 2D and 3D wakes.}
    \label{fig:fig01}
\end{figure}

We consider trajectory planning for wake crossing in 2D and 3D cylinder wakes at $Re=300$ as a canonical problem to examine the effect of three-dimensionality on swimmer navigation.  At this $Re$, the secondary instabilities (referred to as modes A and B) result in three-dimensionality \citep{williamson1996three}. Modes A and B are developed in the vortex cores (seen as less coherent spanwise vortices due to vortex tilting \citep{aleksyuk2023onset} and the braid regions between consecutive spanwise vortices (seen as streamwise-elongated vortices in figure~\ref{fig:fig01}(a)), respectively. These secondary instabilities result in the aperiodic nature of wake flow with a dominant frequency of $St=0.202$, contrary to the periodic flow at a frequency $St=0.212$ developed over a 2D cylinder at the same $Re$. 

We consider wake-crossing scenarios (around 60 samples). These trajectories are shown in red in figure~\ref{fig:fig01}(a). The start (filled circles) and target transverse $y$ locations are on either side of the cylinder wake ($y<0$ for start and $y>0$ for target positions) on the same spanwise plane. The streamwise $x$ locations are chosen randomly in the cylinder wake. The background flow is initialized at the same phase of vortex shedding for both 2D and 3D flows. We consider five different initial phases of vortex shedding as the starting background flow for all the samples. Since the optimization function is performed for each component, the penalization for actuation velocity is also performed component-wise. For these wake-crossing scenarios, the swimmer is sensitive to actuation in a transverse direction, i.e., when not penalized in the transverse direction, the swimmer reaches the target in a straight line, as discussed later. Here, we present the underactuated cases with actuation bounds on the velocity as $0.9,\,0.2,\,0.5$ in the $x,\,y,\,z$ directions with $R_{ii}$ using $(\gamma_1,\,\gamma_2,\,\gamma_3)=(0.1,10,0.1)$.  The time horizon for all cases is chosen as $T_H=10\Delta t\approx0.1T_p$, where $T_p$ is the dominant wake-shedding period. The time horizon can also be measured relative to the size of the vortical structures in the flowfield. The wavelength of the mode B ($\Lambda_B$) instability, resulting in the braid-like region, is $\Lambda_B=0.8D$ \citep{barkley1996three}, hence this time horizon translates to 0.5 convective time units based on the diameter of the cylinder and 0.625 time units relative to the size of the braid-like structures. This indicates that flowfield information about one-half of the size of these vertical structures is enough for successful navigation. The obtained trajectories for $T_H \geq 0.1T_p$ indicate that the swimmers can reach the target when the time horizon is about at least one-tenth of a period. For $T_H=10$, the same trends are observed for $0.85 \leq \eta_1\leq 0.95,\, 0.1\leq \eta_2 \leq 0.3,\, 0.1\leq \eta_2 \leq 0.5$. An instance of the effect of spanwise actuation bounds on the swimmer trajectory is discussed in Appendix (depicted in figure~\ref{fig:rfig01}). Since the optimization strategy minimizes $||e(t)||$ in equation~(2), we consider the swimmer to be successful when $||e(t)||\leq \epsilon D$, which in this study is set to $\epsilon=1/3$.

To compare the navigation performance in 2D and 3D wakes, we use the navigation time $N_T$, the time taken by the swimmer to reach the target, as a performance metric. The probability distribution of $N_T$ for the sampled scenarios averaged over the three different initial background flow conditions is visualized in figure~\ref{fig:fig01}(b). We observe that distribution is shifted toward lower $N_T$ for 3D wake navigation, identifying faster navigation in 3D wakes. The probability distribution also shows cases when $N_T \rightarrow \infty$ for unreachable scenarios, Thus the swimmers navigating in a 3D wake are more likely to reach the target position while being faster than those navigating in a 2D wake.

\begin{figure}[t!]
    \centering
    \includegraphics[scale=0.95]{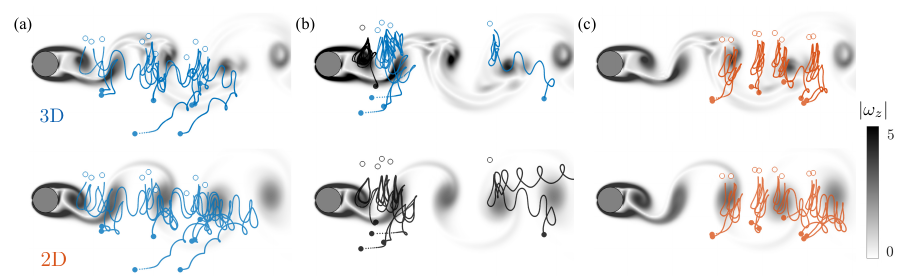}
    \caption{$xy-$ view of wake crossing trajectories in 3D (top row) and 2D (bottom row) wakes visualized using $|\omega_z|$ for when (a) 3D navigation is faster (blue), (b) navigation in 2D cannot reach the target (black) and (c) navigation times in 2D and 3D wakes are similar (red). }
    \label{fig:fig02}
\end{figure}

We divide the obtained trajectories into three scenarios as depicted in figure~\ref{fig:fig02}: (i) when the 3D navigation is faster than the 2D one (blue in figure~\ref{fig:fig02}(a)), (ii) when swimmers cannot reach the target (black in figure~\ref{fig:fig02}(b)) and (iii) when navigation in 2D is faster than in 3D wake (red in figure~\ref{fig:fig02}(c)). For (iii), the navigation time difference is approximately $20\%$ of the wake shedding period $(-\Delta N_T<0.2 T_p)$. The main difference between (i) and (iii) is the initial traverse locations $y_0$ (filled circles): for (i) they are outside of the wake whereas for (iii) the start locations are in the wake region, resulting in lower background velocity from the start. From figures~\ref{fig:fig02}(a), the underactuated swimmer traverses significantly in the streamwise direction before reaching the target, whereas in figure~\ref{fig:fig02}(c) we observe that due to the current actuation parameters and the initial and target locations of the swimmer relative to the cylinder wake, the swimmer can directly cross the wake (almost in a straight line). This shows that streamwise navigation is faster in 3D wakes. 

Figure~\ref{fig:fig02}(b) shows such cases where the swimmer fails to reach the target location in a 2D wake (black) but succeeds in a 3D wake (blue). The transverse target locations $y_{\textrm{target}}>1.4 D$ (unfilled circles) are farther from the wake, located in free-stream, for these cases. The 2D navigation with a finite horizon fails whereas the 3D wake navigation succeeds. Although the target locations are in free-stream, the secondary vortices in the braid regions in 3D wake provide low-velocity regions to reach the target, expanding the reachability boundaries. Even in a 3D wake, there are locations too far from the wake where the swimmer fails to reach the target for a given time horizon (shown by the leftmost trajectory in black in figure~\ref{fig:fig02}(b) (top)).

\begin{figure}
    \centering
    \includegraphics[scale=0.95]{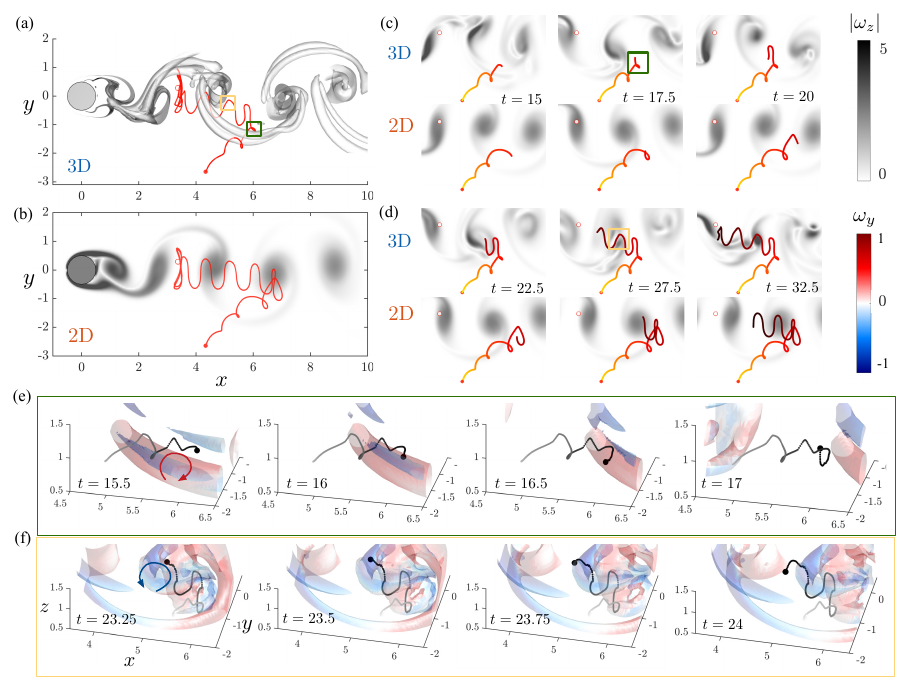}
    \caption{Trajectory optimization for a wake crossing scenario in (a) 3D and (b) 2D wake visualized using $|\omega_z|$. (c)-(d) Instantaneous trajectories during the navigation in 3D (top) and 2D (bottom) wakes. (e)-(f) Zoomed-in view of the evolution swimmer trajectory and background flow visualized using $Q=0.1$ and colored using $\omega_y$ in 3D wake.}
    \label{fig:fig03}
\end{figure}

We now investigate the flow physics that the swimmer leverages for faster navigation in 3D wakes. We consider a scenario where 3D wake navigation is faster ($\Delta N_T>0$) as depicted in figure~\ref{fig:fig03}(a)-(b) with an initial location of $\boldsymbol{x}_0=(4.34,-2.65,1.0)$ (denoted by filled circle) and the target position of $\boldsymbol{x}_{\textrm{target}}=(3.36,0.29,1.0)$ (denoted by unfilled circle) and $\Delta N_T=8.3\approx 1.8T_p$. In both 2D and 3D wake navigation, the swimmer initially travels downstream due to the lower actuation velocity of the swimmer compared to the background flow. A few time instants for the downstream navigation are shown in figure~\ref{fig:fig03}(c). We observe that the swimmer in a 3D wake "redirects" towards the goal earlier than that in a 2D wake as seen at $t=17.5$, even though both the swimmers reach similar $y$ locations. The swimmer in 2D wake travels farther downstream before it can redirect towards the target. Once the $y$ location of the swimmer is in the cylinder wake, utilizing the low- background velocity the swimmer redirects and travels upstream in the wake towards the target as shown in figure~\ref{fig:fig03}(d). The swimmer trajectory oscillates in the $x$ and $y$ directions relative to the spanwise vortices in the cylinder wake, with fewer oscillations observed in 3D wake compared to 2D wake. The upstream navigation trajectory in the 2D wake in figure~\ref{fig:fig03}(b) is similar to the ones identified by \cite{gunnarson2021learning}, where the swimmer leverages the induced velocities in the $x$ and $y$ directions by the spanwise vortices.
 
The zoomed-in view of the time-instants when the swimmer in a 3D wake redirects towards the target (green box in figure~\ref{fig:fig03}(a)) and the time-instants of the swimmer navigating upstream (yellow box in figure~\ref{fig:fig03}(b)) are shown in figures~\ref{fig:fig03}(e)-(f). The instantaneous swimmer location (black dot) and its trajectory (grey) are shown relative to a transverse vortex (shown using $\omega_y$) as it induces velocity in $x$ and $z$ directions. Figure~\ref{fig:fig03}(e) depicts the swimmer navigation under the influence of a counter-clockwise vortex. From, $t=15.5$ to $t=17$, the swimmer first travels in the negative $z$ direction ($t=16,16.5$) and then travels in the upstream (negative $x$) direction, ($t=17$) following the induced velocity and the background flow, thereby redirecting toward the upstream target earlier than its 2D counterpart. 

For efficient upstream navigation in the 3D wakes, two mechanisms are in play: (i) the vorticity in the transverse and streamwise directions and (ii) the reduced coherence of spanwise vortices. Figure~\ref{fig:fig03}(f) depicts the swimmer navigation under the influence of a clockwise transverse vortex while traveling upstream in the wake. From $t=23.25$ to $t=24$, the swimmer first travels in the positive $z$ direction and then in the negative $x$ direction ($t=23.5,23.75$), leveraging the local orientation of background flow caused by the transverse vortex. This shows that the swimmer in a 3D wake travels in positive or negative $z$ direction to effectively leverage the transverse vortices for upstream travel. The swimmer travels in the spanwise direction according to the orientation of subdominant transverse and streamwise vortices. These secondary vortices provide low-velocity regions and change the background flow orientation due to the induced velocity. The swimmer leverages these vortices through the spanwise motion to effectively "redirect" toward the target for faster upstream navigation.

\section{Conclusions}
\label{sec:conclusions}
We investigated the influence of three-dimensionality on the navigation of underactuated swimmers in unsteady wakes using finite-horizon model predictive control for trajectory optimization. We compared the obtained trajectories for navigation in 2D and 3D cylinder wakes at $Re=300$. For both scenarios, the time horizon needed to reach the target is only one-tenth of the wake-shedding period. This makes trajectory optimization for bluff-body wake navigation sensor-friendly. We also identified that the swimmer can navigate faster in 3D wakes compared with 2D wakes for most scenarios. Through spanwise motion, the swimmer can effectively leverage the secondary vortices, specifically transverse vortices to redirect towards the target faster. The low coherence of spanwise vortices and the presence of secondary vortices also facilitate faster upstream navigation when compared with 2D wake navigation.
	
\section*{Acknowledgments}
	\label{sec:acknowledgments}
We acknowledge funding from the US Air Force Office of Scientific Research (FA9550-21-1-0178).  KT thanks the support from the US Department of Defense Vannevar Bush Faculty Fellowship (N00014-22-1-2798). We thank Youngjae Kim for providing the flowfield data and Kai Fukami for his valuable insights.
	
	\section*{Declaration of interest}
	\label{sec:doi}
	The authors report no conflict of interest.

\appendix
\section*{Appendix}
\begin{figure}
        \centering
        \includegraphics[width=0.9\linewidth]{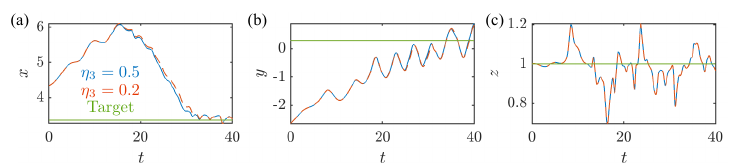}
        \caption{Trajectory evolution for a wake crossing scenario in 3D wake in the three-directions for different actuation bounds in $z$-direction ($\eta_3$).}
        \label{fig:rfig01}
    \end{figure}

{\label{sec:appendix}}
The underactuated swimmer trajectories shown in the main manuscript are obtained using the actuation bounds of $(\eta_1,\eta_2,\eta_3)=(0.9,0.2,0.5)$ in $x,y$ and $z$ directions respectively. While the lower actuation bound in the $y$ direction is needed to avoid crossing the wake in a straight line, we here address the effect of actuation bounds in the spanwise direction. Let us consider the scenario discussed in figure~\ref{fig:fig03}. Figure~\ref{fig:rfig01} shows the trajectory evolution of the swimmer when $\eta_3=0.2$ (red) and $\eta_3=0.5$ (blue). Here, we use the same actuation bounds of $(\eta_1,\eta_2)=(0.9,0.2)$ in $x$ and $y$ directions. We observe that both trajectories are similar irrespective of the difference in the actuation bound in the $z$ direction. Since the secondary vortices are weaker than the spanwise vortices, the induced velocity by the secondary vortices in the $z$ direction is also quite small, resulting in less actuation requirement in the $z$ direction.

 \bibliographystyle{unsrtnat}
 \bibliography{refs}

\begin{thebibliography}{21}
\providecommand{\natexlab}[1]{#1}
\providecommand{\url}[1]{\texttt{#1}}
\expandafter\ifx\csname urlstyle\endcsname\relax
  \providecommand{\doi}[1]{doi: #1}\else
  \providecommand{\doi}{doi: \begingroup \urlstyle{rm}\Url}\fi

\bibitem[Watkins et~al.(2020)Watkins, Burry, Mohamed, Marino, Prudden, Fisher, Kloet, and Clothier]{watkins2020ten}
S.~Watkins, J.~Burry, A.~Mohamed, M.~Marino, S.~Prudden, A.~Fisher, N.~Kloet, and T.~Jakobi {\&}~R. Clothier.
\newblock Ten questions concerning the use of drones in urban environments.
\newblock \emph{Build. Environ.}, 167:\penalty0 106458, 2020.

\bibitem[Shukla and Singh(2019)]{shukla2019ship}
S.~Shukla and S.~S. Sinha {\&} S.~N. Singh.
\newblock Ship-helo coupled airwake aerodynamics: a comprehensive review.
\newblock \emph{Prog. Aerosp. Sci.}, 106:\penalty0 71--107, 2019.

\bibitem[Zereik et~al.(2018)Zereik, Bibuli, {s}kovi\'{c}, and Pascoal]{zereik2018challenges}
E.~Zereik, M.~Bibuli, N.~Mi\ {s}kovi\'{c}, and P.~Ridao {\&}~A. Pascoal.
\newblock Challenges and future trends in marine robotics.
\newblock \emph{Annu. Rev. Control}, 46:\penalty0 350--368, 2018.

\bibitem[Ortega-Jimenez et~al.(2014)Ortega-Jimenez, Sapir, Wolf, and Dudley]{ortega2014into}
V.~M. Ortega-Jimenez, N.~Sapir, M.~Wolf, and E.~A. Variano {\&}~R. Dudley.
\newblock Into turbulent air: size-dependent effects of von k{\'a}rm{\'a}n vortex streets on hummingbird flight kinematics and energetics.
\newblock \emph{Proc. Roy. Soc. B}, 281\penalty0 (1783):\penalty0 20140180, 2014.

\bibitem[Shyy et~al.(2016)Shyy, Kang, Chirarattananon, and Liu]{shyy2016aerodynamics}
W.~Shyy, C.~K. Kang, P.~Chirarattananon, and S.~Ravi {\&}~H. Liu.
\newblock Aerodynamics, sensing and control of insect-scale flapping-wing flight.
\newblock \emph{Proc. Roy. Soc. A}, 472\penalty0 (2186):\penalty0 20150712, 2016.

\bibitem[Liao(2007)]{liao2007review}
J.~C. Liao.
\newblock A review of fish swimming mechanics and behaviour in altered flows.
\newblock \emph{Phil. Trans. Roy. Soc. B}, 362\penalty0 (1487):\penalty0 1973--1993, 2007.

\bibitem[Lauder(2015)]{lauder2015fish}
G.~V. Lauder.
\newblock Fish locomotion: recent advances and new directions.
\newblock \emph{Annu. Rev. of Mar. Sci.}, 7\penalty0 (1):\penalty0 521--545, 2015.

\bibitem[Maia and Tytell(2015)]{maia2015streamwise}
A.~Maia and A.~P. Sheltzer {\&} E.~D. Tytell.
\newblock Streamwise vortices destabilize swimming bluegill sunfish (lepomis macrochirus).
\newblock \emph{J. Exp. Bio.}, 218\penalty0 (5):\penalty0 786--792, 2015.

\bibitem[Kularatne and Hsieh(2016)]{kularatne2016time}
D.~Kularatne and S.~Bhattacharya {\&} M.~A. Hsieh.
\newblock Time and energy optimal path planning in general flows.
\newblock In \emph{Robotics: Science and Systems}, pages 1--10. Ann Arbor, MI, 2016.

\bibitem[Lermusiaux(2016)]{subramani2016energy}
D.~N. Subramani {\&} P.~F. Lermusiaux.
\newblock Energy-optimal path planning by stochastic dynamically orthogonal level-set optimization.
\newblock \emph{Ocean Model.}, 100:\penalty0 57--77, 2016.

\bibitem[Colabrese et~al.(2017)Colabrese, Gustavsson, Celani, and Biferale]{colabrese2017flow}
S.~Colabrese, K.~Gustavsson, A.~Celani, and L.~Biferale.
\newblock Flow navigation by smart microswimmers via reinforcement learning.
\newblock \emph{Phys. Rev. Lett.}, 118\penalty0 (15):\penalty0 158004, 2017.

\bibitem[Gunnarson et~al.(2021)Gunnarson, Mandralis, Novati, and Dabiri]{gunnarson2021learning}
P.~Gunnarson, I.~Mandralis, G.~Novati, and P.~Koumoutsakos {\&} J.~O. Dabiri.
\newblock Learning efficient navigation in vortical flow fields.
\newblock \emph{Nat. Commu.}, 12\penalty0 (1):\penalty0 7143, 2021.

\bibitem[Jiao et~al.(2021)Jiao, Ling, Heydari, Heess, and Kanso]{jiao2021learning}
Y.~Jiao, F.~Ling, S.~Heydari, N.~Heess, and J.~Merel {\&}~E. Kanso.
\newblock Learning to swim in potential flow.
\newblock \emph{Phys. Rev. Fluids}, 6\penalty0 (5):\penalty0 050505, 2021.

\bibitem[Ross(2008)]{senatore2008fuel}
C.~Senatore {\&} S.~D. Ross.
\newblock Fuel-efficient navigation in complex flows.
\newblock In \emph{American Control Conference}, pages 1244--1248. IEEE, 2008.

\bibitem[Krishna and Brunton(2022)]{krishna2022finite}
K.~Krishna and Z.~Song {\&} S.~L. Brunton.
\newblock Finite-horizon, energy-efficient trajectories in unsteady flows.
\newblock \emph{Proc. Roy. Soc. A}, 478\penalty0 (2258):\penalty0 20210255, 2022.

\bibitem[Iaccarino(2004)]{ham2004energy}
F.~Ham {\&}~G. Iaccarino.
\newblock Energy conservation in collocated discretization schemes on unstructured meshes.
\newblock \emph{Annual Research Briefs}, 2004\penalty0 (3-14):\penalty0 118, 2004.

\bibitem[Ham and Iaccarino(2006)]{ham2006accurate}
F.~Ham and K.~Mattson {\&}~G. Iaccarino.
\newblock Accurate and stable finite volume operators for unstructured flow solvers.
\newblock \emph{Annual Research Briefs}, 243, 2006.

\bibitem[Kim et~al.(2024)Kim, Godavarthi, Rolandi, and Taira]{kim2024influence}
Y.~Kim, V.~Godavarthi, V.~Rolandi, and J.~Klamo {\&}~K. Taira.
\newblock Influence of three-dimensionality on wake synchronisation of an oscillatory cylinder.
\newblock \emph{J. Fluid Mech.}, 1001:\penalty0 A24, 2024.

\bibitem[Williamson(1996)]{williamson1996three}
C.~H.~K Williamson.
\newblock Three-dimensional wake transition.
\newblock \emph{J. Fluid Mech.}, 328:\penalty0 345--407, 1996.

\bibitem[Aleksyuk and Heil(2023)]{aleksyuk2023onset}
A.~I. Aleksyuk and M.~Heil.
\newblock On the onset of long-wavelength three-dimensional instability in the cylinder wake.
\newblock \emph{J. Fluid Mech.}, 967:\penalty0 A23, 2023.

\bibitem[Barkley and Henderson(1996)]{barkley1996three}
D.~Barkley and R.~D. Henderson.
\newblock Three-dimensional floquet stability analysis of the wake of a circular cylinder.
\newblock \emph{J. Fluid Mech.}, 322:\penalty0 215--241, 1996.

\end{thebibliography}

\end{document}